\input harvmac

\def\a{\alpha}

\def\d{\lambda}
\def\g{\gamma}
\def\eps{\epsilon}

\def\p{\negthinspace \negthinspace}
\def\intl{\int\limits_{-\infty}^{+\infty}}
\def\nl{\intl dx\intl dx'\intl dx''}
\def\l{\langle}
\def\r{\rangle}
\def\ra{\rightarrow}
\def\q{\vert \s (0)\vert^2}

\def\o{\over}
\def\s{\Psi}
\def\t{\theta}
\def\u{\uppercase\expandafter}
\def\y{\sim}
\def\({\left (}
\def\){\right )}
\def\[{\left [}
\def\]{\right ]}
\def\Td{T_{\rm decoherence}}

\lref\Harb{J. B. Hartle, Phys. Rev. D {\bf 44}, 3173 (1991)} 
\lref\Feyn{R.P.~Feynman, Rev. Mod. Phys., 20, 267, 1948.}
\lref\Mensky{M.B.~Mensky, Phys. Rev. D, 20, 384, 1979;
Theor. Math. Phys., 75, 357, 1988.}
\lref\Caves{C.~Caves,  Phys. Rev. D, 33, 1643, 1986; {\it ibid.},
  35, 1815, 1987.}
\lref\Schmid{A.~Schmid, Ann. Phys., 173, 103, 1987.}
\lref\Har{J.B.~Hartle, Phys. Rev. D, 38, 2985, 1988.}
\lref\Mar{D. Marolf, Phys. Rev. A, 50, 939, 1994}
\lref\YTa{N.~Yamada and S.~Takagi, Prog. Theor. Phys., 85,
985, 1991.}
\lref\Sor{R.~Sorkin, Int. Jour. Theor. Phys.,
33, 523, 1994.}
\lref\Hara{J.B.~Hartle, {\it The Quantum Mechanics of Cosmology}, in {\sl
Quantum Cosmology and Baby Universes:  Proceedings of the 1989 Jerusalem 
Winter
School for Theoretical Physics}, ed. by ~S.~Coleman, J.B.~Hartle, T.~Piran,
and S.~Weinberg, World Scientific, Singapore (1991) pp. 65-157.}
\lref\YTb{N.~Yamada and S.~Takagi, Prog. Theor. Phys., 86,
599, 1991; {\it ibid.}, 87, 77, 1992.}
\lref\GR{I.S.~Gradshteyn and I.M.~Ryzhik, {\sl Table of Integrals, Series,
and Products}, (Academic Press, New York, 1965)}

\vskip-.2in
\line{\hfil NSF--ITP--96--13}
\vskip-.7in
\Title{\vbox{\baselineskip12pt \hbox{ }}}
{\vbox{\centerline {Nearly Instantaneous Alternatives in  
Quantum Mechanics}}}
\vskip-.1in
\centerline{Richard J. Micanek}
\centerline{\sl Department of Physics}
\centerline{\sl University of California}
\centerline{\sl Santa Barbara, CA 93106-9530}
\centerline{\sl micanek@vorpal.ucsb.edu}
\vskip.1in
\centerline{James B. Hartle}
\centerline{\sl Institute for Theoretical Physics}
\centerline{\sl University of California}
\centerline{\sl Santa Barbara, CA 93106-9530}
\centerline{\sl hartle@itp.ucsb.edu}

\bigskip
\centerline{\bf Abstract}

Usual quantum mechanics predicts probabilities for 
the outcomes of measurements carried out at definite moments of time. 
However, realistic measurements do not take place in an instant, but are 
extended over a period of time.  The assumption of instantaneous 
alternatives in usual quantum mechanics is an approximation 
whose validity can be investigated in the generalized quantum mechanics
of closed systems in which probabilities are predicted for spacetime
alternatives that extend over time.  In this paper we investigate how
alternatives extended over 
time reduce to the usual instantaneous alternatives in a simple 
model in non-relativistic quantum mechanics. Specifically, we show how
the decoherence of a particular set of spacetime alternatives
becomes automatic
as the time over which they extend approaches zero and estimate
how large this time can be before the interference between the
alternatives becomes non-negligible.  
These results suggest that the time scale over which coarse
grainings of such quantities as the center of mass position of a massive
body may be \nobreak extended in time before producing significant interference is much longer than characteristic dynamical time scales. 

\vskip.05in
\Date{2/96}

\bigskip
\centerline{\bf I. INTRODUCTION}

As usually formulated, quantum mechanics predicts probabilities for 
the outcomes of measurements carried out at definite moments of time. 
However, realistic measurements do not take place in an instant, but are 
extended over a period of time.  The assumption of instantaneous 
measurements in usual quantum mechanics is an approximation 
whose validity can be 
investigated in the generalized quantum mechanics of closed systems 
where probabilities are predicted for alternatives that are extended 
over time.  In this paper we investigate how alternatives extended over 
time reduce to the usual alternatives at a moment of time in a simple 
model in non-relativistic quantum mechanics.

There are a number of interesting discussions of ideal measurements 
extended over time in the quantum mechanical literature 
\refs{\Feyn,\Mensky,\Caves,\Schmid,\Har,\YTa,\Sor,\Mar}.
However, these 
discussions were incomplete because they did not completely specify what 
an ``ideal" measurement consisted of nor what replaced the reduction of 
the state vector upon the completion of one.  In the more general quantum 
mechanics of closed systems, in which a notion of ``measurement" 
does not play a fundamental role, clear meaning can be given to the 
probabilities for spacetime alternatives which extend over time 
\refs{\Hara,\Harb,\YTb}. 

Spacetime alternatives are easily visualized in a sum-over-histories 
formulation of quantum mechanics. Consider for example the quantum 
mechanics of a single non-relativistic particle moving in one spatial 
dimension. The {\it fine-grained-histories} for this system are the 
particle paths $x(t)$ on a time interval, say,  $[0,T]$.  The most general sets
of alternatives for this system are partitions of this set of 
fine-grained paths into sets of coarse-grained classes, $c_{\alpha}$, 
where $\alpha = 1,2,3\cdots$.  Partitions by the values of $x$ at which 
paths cross a surface of time $t$ are  simple kinds of coarse grainings that
correspond to the usual alternatives of position at a moment of time. Partitions into 
classes whose definition extends over time generalize these to spacetime 
alternatives. A simple example is a partition of the paths defined by 
whether they cross or never cross a given spacetime region with extent in time.

For each class of paths $c_{\alpha}$, a class operator $C_{\alpha}$ may be 
defined by a path integral over paths in the class of the form
\eqn\one{
\langle x''|C_{\alpha}|x'\rangle = \int _{\alpha}\delta x 
{\rm exp}\left( iS[x(\tau)]/ \hbar\right).}
Here, $S[x(\tau)]$ is the action functional for paths and the sum is over
all paths in the class $c_\alpha$. This class operator 
incorporates both unitary dynamics and a generalization of state vector 
reduction in a unified way. When the system's initial state at $t=0$ is 
$|\Psi\rangle$, the probability of coarse-grained alternative $\alpha$ is 
\eqn\two{
p(\alpha) = \| C_{\alpha}|\Psi\rangle\|^2.}
However, the quantum mechanics of closed systems does not predict 
probabilities for every set of coarse-grained alternatives that may be 
described, but only for those which have negligible quantum 
mechanical interference between the individual histories in the set.  
Such sets are said to decohere. Specifically, sets of histories for which 
consistent probabilities are predicted must satisfy a decoherence 
condition, which for present purposes we may take to be 
\eqn\three{
\langle\Psi|C^{\dagger}_{\a'}\cdot C_{\a}|\Psi\rangle \sim 
0 \ ,\quad \a'\neq\a \ .}
All the required probability sum rules are satisfied by the probabilities 
defined by (2) as a consequence of the decoherence condition (3).

It is a straightforward calculation \Harb\ to verify that alternatives 
defined at one moment of time $t$ have class operators of the form 
\eqn\four{
C_{\alpha} = e^{-iHT}P_{\alpha}(t)}
where $H$ is the Hamiltonian of the closed system and $\{ P_{\alpha}(t)\}$ 
are a set of orthogonal Heisenberg picture projection operators. 
(Here, as throughout, we use units in which $\hbar=1$.)
For instance, for the alternatives that the particle is in one of a set 
of exclusive spatial regions $\{ \Delta_{\alpha}\}$ the $P$'s are the 
projections onto these regions at the appropriate time. It is then
immediate from (3) that decoherence is automatic for such instantaneous
alternatives. 

Decoherence is not automatic for alternatives that are extended over 
time. However as the time $T$ over which they extend approaches 
zero they must become decoherent and their class operators must approach 
the form (4). This paper examines this approach of spacetime alternatives 
to instantaneous ones 
in a simple model. The model is described in Section II and 
its behavior for small $T$ found in Sections III -- V. The significance
of the results is discussed in Section VI.

\bigskip
\centerline{\bf II. A MODEL COARSE-GRAINING}

Our model concerns a non-relativistic particle of mass $M$, moving in
one dimension, in a potential $V(x)$.
The fine-grained histories for this system between times $t=0$ and
$t=T$ are particle paths represented by single valued functions $x(t)$
on that interval. Coarse-grainings are generally partitions of these
paths into sets of exclusive classes. The individual classes are
called coarse-grained histories.

As a simple example of a spacetime coarse graining that extends over time,
consider the region  
of spacetime $R$ that lies between times $t=0$ and $t=T$.  
Denote the subregions of $R$ that lie to the left and to the  
right of the origin as $R_l$ and $R_r$, respectively. The set of all
paths between $t=0$ and $t=T$ may be partitioned by whether they
cross or do not cross the left and right regions $R_l$ and $R_r$. 
Specifically, let  
$c_{10}$ be the class of paths which  
remain to the left of the origin and never cross $R_r$,
let $c_{01}$ be the class of  
paths that stay to the right of the origin and never cross $R_l$, and let
 $c_{11}$ be the class of paths  that cross both regions sometime.
The class $c_{00}$ of paths that never  
cross either $R_l$ or $R_r$ is empty and we shall not discuss it further. 
This set of  
coarse-grained histories provides a simple model for  
investigating the limit of small temporal extension 
$T$. In that limit, the set of spacetime alternatives should approximately 
decohere. Further, the probability for the alternative $c_{01}$ in which the
particle is localized on the right during time $T$ should approach
the usual probability for the particle's position at $t=0$ to be  located 
in the region $x>0$.  There should be a similar approach of the probability 
for $c_{10}$ to the usual probability that the particle is located in $x<0$ at
$t=0$. The probability of the alternative $c_{11}$ that the particle is in 
{\it both} regions of $x$ should approach zero. 
In the following we shall show that these expectations are correct.

As we shall show in Section V, a bounded potential $V(x)$ has a negligible
effect on the class operators for very short times $T$. We therefore
begin with an investigation of their form for a free particle with
$V(x)=0$.
The class operators for the model coarse-graining  were calculated 
in \Harb but we briefly review their construction here. They may 
all be expressed
in terms of the free particle propagator for the time interval $T$
which is
\eqn\prop{K_T (x'',x') = \(\d \o {i\pi} \)^{1 \over 2} e^{i\d  
(x''-x')^2} ,}
where 
\eqn\lam{\d = {M \o {2T}}}
is a parameter which becomes large as $T$ becomes small. 
Consider for example the operator 
$C_{01}$ corresponding to the class  of paths that remain entirely
to the right of $x=0$ for the time $T$. Its matrix elements are
given by path integrals of the form \one\ over this class of paths. 
That path integral is the same as the path integral over {\it all}
paths with an action including an infinite barrier potential for $x<0$. That is, 
the class operator is the ordinary quantum mechanical propagator in the presence
of an infinite reflecting  barrier at $x=0$. The appropriate solution
of the Schr\"odinger equation may be found by the method of images and
is 
\eqn\meli{ \l x'' \vert C_{01} \vert x' \r = \t(x'')\t(x') \[  
K_T(x'',x') - K_T(-x'',x')\] \ . }
Similarly, 
\eqn\melii{ \l x'' \vert C_{10} \vert x' \r = \t(-x'')\t(-x')  
\[ K_T(x'',x') - K_T(-x'',x')\] \ . }
To find the remaining class operator $C_{11}$ note that a sum of the form \one\
over {\it all} paths just gives the usual
free particle propagator, so that 
\eqn\som{\sum_\a C_\a = e^{-iHT} \ . }
Thus 
\eqn\meliii{\eqalign{ \l x'' \vert C_{11} \vert x' \r & =   
K_T(x'',x') - \l x'' \vert C_{01} \vert x' \r -  \l x'' \vert  
C_{10} \vert x' \r \cr & = \[ \t(x'')\t(-x') + \t(-x'')\t(x')  
\] K_T(x'',x') \cr & \qquad + \[ \t(x'')\t(x') + \t(-x'')\t(-x')  
\] K_T(-x'',x') \ . \cr}}
With these class operators the decoherence functional for this  
set of coarse-grained histories may be computed in the limit  
of vanishing temporal $T$.

\bigskip
\centerline{\bf III. THE DECOHERENCE FUNCTIONAL}

We now turn to the question of the decoherence and probabilities of 
the set of spacetime alternatives described above as a function
of the time $T$ over which they extend. For simplicity, we 
assume that at $t=0$ the particle is in a pure state represented by
a wave function $\Psi(x)$. The decoherence functional defined
by 
\eqn\fund{D(c_\a,c_{\a'}) = \langle\Psi| C^{\dag}_{\a'}C_{\a}|\Psi\rangle} 
is a convenient tool for summarizing the essential features of
the quantum mechanics of closed systems that were mentioned in the Introduction.
A set of spacetime alternatives decoheres when the off-diagonal
elements of $D$ are negligibly small; the probabilities of a decoherent set
of alternatives are given by the diagonal elements of $D$.

The decoherence functional \fund\ may be expressed in terms
of the matrix elements of the class operators found in the preceding
Section by writing
\eqn\states{D(c_\a,c_{\a'}) = \nl \s(x') \s^*(x'')
\l x \vert  C_\a \vert x' \r \l x \vert C_{\a'} \vert x'' \r^*  \ . }
Given that $c_{00}=0$ is empty,  it is clear that any element of $D$  
involving this class vanishes identically. 
Additionally, since $\t(x)\t(-x)=0$ we have 
\eqn\thota{D(c_{01},c_{10}) = D^*(c_{10},c_{01}) = 0 \ . }
The hermiticity of the decoherence functional implies the additional  
relations $D(c_{01},c_{11}) = D^*(c_{11},c_{01})$ and  
$D(c_{10},c_{11}) = D^*(c_{11},c_{10})$, leaving only five  
components to compute. We now describe how to carry out the 
integrals in \states\ for these components.

We begin with the diagonal element $D(c_{01},c_{01})$. 
It follows from \fund\ that this
element of the decoherence functional is the square of the norm
of the vector $C_{01}|\Psi\rangle$. The class operator $C_{01}$
given by \meli\ is  the propagator for Schr\"odinger evolution over
time $T$ of a free particle in the presence of an infinite barrier on 
$x<0$. Applied to $|\Psi\rangle$ it gives the  {\it branch} wave function 
$\Psi_{01}(x)=\langle|C_{01}|\Psi\rangle$ with
\eqn\branch{\s_{01} (x,T) \equiv \int\limits_0^\infty dx' \[K_T(x,x')-K_T(x,-x')\] \Psi(x') = \int\limits_{- \infty}^{+ \infty} dx' K_T(x,x') {\widetilde\Psi}(x') \ ,}
where 
\eqn\disc{\widetilde\Psi(x) = \Psi(x) \ ,  x>0 \qquad  {\rm and}  \qquad
\widetilde\Psi(x)= -\Psi(-x), \ x<0.  \qquad}
This is similar to evolution of the usual Schr\"odinger equation but with discontinuous initial data.  Notice that the probability integrals are unaffected because this data resides in the Hilbert space $L_2$.  Thus the norm evaluated on the interval 
$x\ge 0$ is preserved under this evolution, so that
\eqn\pb{D(c_{01},c_{01}) = \int\limits_0^{\infty} dx |\Psi(x)|^2 \equiv p_+}
and is {\it independent} of the time interval $T$. Similarly,
\eqn\pba{D\(c_{10},c_{10}\) = \int\limits_{-\infty}^0 dx |\Psi(x)|^2 \equiv p_- \ .}
When the set of spacetime alternatives decoheres, the diagonal element
$D(c_{01}, c_{01})$ is the probability that the particle remains at positive
values of $x$ throughout the time interval $T$.  For this model this
is the same as the probability that the particle is in this range of
$x$ at $t=0$.

The remaining elements of the decoherence functional, which we expect
to vanish in the limit of vanishing $T$, may be evaluated as follows:
Consider as a typical example the interference term $D(c_{01},c_{11})$.
Using the results of Section II for the matrix elements of the
class operators, the integrations in \states\ can be rearranged to
give
\eqn\X{D(c_{01},c_{11}) = \p \int\limits_0^\infty \p dx \p \int\limits_0^\infty \p dx' \p 
\int\limits_0^\infty \p dx'' \ K^*_T(x,-x'')  \[K_T(x,x')-K_T(x,-x')\]\Psi_S^*(x'')
\Psi(x')}
where
\eqn\Y{\Psi_{S}(x)=\Psi(x)+\Psi(-x) \ .}

It is convenient to rescale the integration variables by writing
$x=y/\lambda^{1/2}$ with similar rescalings for $x'$ and $x''$. Then,
using the explicit form \prop\ for the free particle 
propagator $K_T(x'',x')$, we have 
\eqn\Z{\eqalign{D(c_{01},c_{11}) = & \ {1 \o {\pi\lambda^{1/2}}}  
\int\limits_0^\infty dy  \int\limits_0^\infty  dy'  
\int\limits_0^\infty  dy'' e^{-i(y+y'')^2}[e^{i(y-y')^2}-e^{i(y+y')^2}]
\cr & \qquad \times \Psi_S^* \( {y'' \o \lambda^{1/2}} \) \Psi \( {y'\o \lambda^{1/2}} \). \cr}}
{F}rom this expression it is clear that only the behavior of $\Psi(x)$ near
$x=0$ will contribute in the limit of large $\lambda$ and small $T$.
Inserting a factor of $e^{-\epsilon(y'^2+y''^2)}$ to ensure the convergence
of the integral, the expression has the following small $T$ asymptotic
form for finite $\epsilon$:
\eqn\malt{D_{\eps}(c_{01},c_{11}) \y {4 \o {i\pi\lambda^{1/2}}} 
\q \int\limits_0^\infty  
dy' e^{- \xi^* y{'^2}} \int\limits_0^\infty dy'' e^{-\xi y{''^2}}  
\int\limits_0^\infty dy \ e^{-2iy''y} \sin (2y'y)}
where 
\eqn\epsi{\xi= \epsilon +i \ .}

Using eqs. $(3.893.1)$, $(3.466.1)$ and $(6.286.1)$ of \GR\ ,    
the integral in \malt\ may be evaluated explicitly in terms
of a hypergeometric function:
\eqn\hyper{{1 \o 4} \sqrt{\pi \o \xi} {_2F_1}({1 \o 2},1;{3 \o  
2};{{2\eps } \o \xi}) \ . }
Now insert this in \malt\ and pass to the $\eps \ra 0$ limit to remove 
the convergence  factor and find the simple result:
\eqn\final{D(c_{01},c_{11}) \y -{{e^{{i\pi} \o 4}} \o  
\sqrt{\pi \d}} \q \ . }

No additional calculation is needed to evaluate the remaining 
off-diagonal component $D(c_{10},c_{11})$. That is because, using
the class operator matrix elements from Section II and changing
the sign of the integration variables $x$, $x'$ and $x''$, one obtains 
an expression for $D(c_{10},c_{11})$ which is identical to
the right hand side of \X\ except with $\Psi(x')$ replaced with
$\Psi(-x')$. However, in the limit of small $T$,
only the value
$\Psi(0)$ is important as \malt\ shows. The leading orders of 
$D(c_{10},c_{11})$ and $D(c_{01},c_{11})$ therefore coincide 
for small $T$, both given by the right hand side of \final.

The remaining element of the decoherence functional is 
$D(c_{11},c_{11})$.  This can be evaluated using the techniques
employed above for the other elements which vanish in the limit of  small $T$.
However, it is quicker simply to evaluate it from the general
relation
\eqn\form{\sum_{\a,\a'} D(c_\a,c_{\a'}) = 1}
which follows from \som\ and \fund\ .
Either way  the result is:
\eqn\dii{D(c_{11},c_{11}) \y 2 \sqrt{2 \o {\pi \d}} \q}
to leading order in $\d$.

Putting these results together, the small $T$ behavior of the
decoherence functional for the three nontrivial alternatives is given by
\eqn\matrl{D(c_\a,c_{\a'}) = \pmatrix {p_+ & 0 & 0 \cr 0 & p_- & 0 \cr 
0 & 0 & 0 \cr} + \pmatrix {0 & 0 & \eta \cr 0 & 0 & \eta \cr \eta^* & \eta^* & -2 \[ \eta + \eta^* \]} + \cdots}
with $p_\pm$ as in \pb\ and \pba\ and
\eqn\neta{\eta = -{{e^{{i\pi} \o 4}} \o \sqrt{\pi \d}} \q \ ,}
where row and column are taken in the order $c_{01}, c_{10}, c_{11}$.
The following points are immediate consequences of this expression:
\item{}(1) The off-diagonal elements of the decoherence functional
vanish as $T^{1/2}$ in the limit of small $T$
so that this set of spacetime alternatives decoheres in that limit.
\item{}(2) In the limit of vanishing $T$,  non-vanishing diagonal elements 
of the decoherence functional coincide exactly with
the probabilities for the particle to be on the left or the right of
$x=0$ at the moment of time $t=0$, as expected.
\item{}(3) Unlike the case of histories which are sequences of 
sets of alternative projections, the diagonal elements of the
decoherence functional for spacetime alternatives are not probabilities
when the set of alternatives are not decoherent. They do not sum to
one. (See  \Harb\ for a more general discussion.)
\item{} (4) The leading order of the interference terms vanishes if
$\Psi(0)=0$. That is consistent with the results of \YTb\  who 
showed that decoherence is exact for any time interval $T$ as long
as the initial wave function $\Psi(x)$ is antisymmetric about
$x=0$ (a result which follows immediately from \X\ since $\Psi_S(x)$ 
then vanishes).

\bigskip
\centerline{\bf IV. A SPECIFIC INITIAL CONDITION}

An initial Gaussian wave packet provides an example in which we can 
explicitly evaluate the decoherence functional for the spacetime 
coarse graining under discussion without
recourse to the limit of small $T$. Assume a 
one-dimensional Gaussian initial wave packet 
for our free particle of the form
\eqn\gauss{\s (x) = \(2 \o {\pi \ell^2}\)^{1 \o 4} e^{-x^2 /  
\ell^2}}
where $\ell$ is the characteristic width of the wave packet.  
With \gauss\ into \states, evaluation of the relevant integrals that are 
simply those of the preceding Section yields
\eqn\matr{\eqalign{D(c_\a,c_{\a'}) = \pmatrix {{1 \o 2} & 0 & 0 \cr
0 & {1 \o 2} & 0 \cr 0 & 0 & 0 \cr} + \pmatrix {0 & 0 & \g \cr 
0 & 0 & \g \cr \g^* & \g^* & -2\[ \g + \g^* \] \cr}}}
as the desired decoherence matrix with
\eqn\gamm{{\g = {1 \o {i\pi}}{\rm arctanh}^{-1} \(\sqrt{2 \o {1 +  
i\ell^2 \d}} \) = \eta + \cdots} \ ,}
having put $\s (0) = \(2 \o {\pi \ell^2}\)^{1 \o 4}$ in \neta\ for this special
case.  As expected, in the limit of large $\lambda$ or small $T$,  
the results of \matrl\ are reproduced and decoherence is
achieved with 
equal probabilities of $1/2$ for each of the non-vanishing  
coarse-grained alternatives.
\eject

\bigskip
\centerline{\bf V. INCLUDING A POTENTIAL}
We now consider the effect of  including a bounded potential on the 
above results
derived for a free particle. Classically, the effect of a bounded force
on the motion of a particle is proportional to the time interval
over which it acts as long as that time interval is sufficiently small.
The effect of a potential is 
thus negligible for very small time intervals. We expect
a similarly negligible effect of a bounded potential on the propagation
of a particle in quantum mechanics when the time of propagation becomes
small (as it does in the coarse grainings under discussion in this paper).
We shall now show that this is the case.

The general arguments involving the conservation of probability
that led to the the results \pb\ and \pba\ for the diagonal
elements of the decoherence functional, $D(c_{01},c_{01})$ and 
$D(c_{10},c_{10})$, respectively,
are as valid in the presence of a potential $V(x)$ as they were without.
The values of these elements of $D$ are therefore unchanged by the inclusion
of a potential for any value of $T$.

To understand the effect of a potential on the other elements of the decoherence
functional which vanish in the limit of small $T$, we need to examine
the propagator over a time interval $T$. In the presence of a 
potential this is:
\eqn\Pr{{G_T(x'',x')=\langle x''|e^{-i({ H}_0 +  V)T}|x' \rangle} \ ,}
where ${ H}_0$ is the free particle Hamiltonian. It is not difficult to see that
introducing a potential changes the construction of the class operator
matrix elements in Section II only by replacing the free particle
propagator $K_T(x'', x')$ with $G_T(x'',x')$. 
The results for the decoherence functional can then all be expressed in 
terms of inner products of wave functions of the form
\eqn\Wf{\int dx' G_T(x,x')\Psi(x') \ .}
We now find explicit expressions for the modifications induced by the 
potential on wave functions of this form.

The evolution operator $U(t) \equiv {\rm exp}[-i({ H}_0 +  V )t]$
satisfies the Schr\"odinger equation
\eqn\Seq{ id U(t) /dt = ({ H}_0 +  V )  U(t) \ .}
The correct solution can be written in
the form
\eqn\epn{U(t)=e^{-i { H}_0 t}[1-i\int\limits_0^t dt' e^{i{ H}_0 t'}
{ V} e^{-i{ H}_0 t'} + \cdots ] \ .}
Since the free particle propagator is 
\eqn\fp{K_T(x'',x')=\langle x''|e^{-i{ H}_0 T}|x' \rangle \ ,}
this result can be used to write the evolution of an initial wave 
function $\Psi(x)$ over a time interval $T$ in the form
\eqn\eqv{\int\limits_{-\infty}^{+\infty} dx' G_T(x,x')\Psi(x') = \int
\limits_{-\infty}^{+\infty} dx' K_T(x,x')\Psi^V_T(x')}
where
\eqn\pv{\Psi^V_T(x)=[1-i\int\limits_0^T dt' e^{i{ H}_0 t'}
 V e^{-i{ H}_0 t'} + \cdots ]\Psi(x) \ .}
The calculations of the small time behavior of the decoherence 
functional are thus the same as those in Section III with $\Psi^V_T(x)$
replacing $\Psi(x)$. However, since we are interested 
only in the leading order in small $T$, we may employ only the leading
order in $\Psi_T^V(x)$.  That just comes from the first
term in the sum in \pv, the higher order terms in the potential vanishing
as successively higher order powers of $T$. The leading term is thus 
just $\Psi(x)$.
Including a potential $V(x)$ therefore does not change the small $T$
asymptotic form of the decoherence functional given by \matrl.

\bigskip
\centerline{\bf VI. DISCUSSION}

We have investigated a very simple example of a spacetime coarse
graining in the quantum mechanics of a single non-relativistic
particle of mass $M$ moving in one dimension.  The three
non-empty coarse-grained 
alternatives are  whether the particle
remains always to the right of $x=0$ for a time interval $T$,
remains always to the left of $x=0$ for this interval, or is
sometimes on the left and sometimes on the right during that time. 
As the example discussed in the previous Section shows, when the
initial state is a wave packet of width $\ell$, the
characteristic time scale for the automatic decoherence of 
this set of alternatives is $T_{\rm decoherence}  \y {{M\ell^2}/\hbar}$. 
When $T\ll T_{\rm decoherence}$, the interference between these alternatives
is negligible and the set approximately decoheres. The exact decoherence
of instantaneous alternatives is thus a good approximation to the
nearly exact decoherence of this type of spacetime alternative.  
Further, the probabilities of the instantaneous alternatives are
exactly the same as those of the spacetime ones. When $T\ll \Td$ 
instantaneous alternatives are an excellent approximation to this
kind of spacetime alternative.

To get a feel for this scale, consider an electron localized to its  
Compton wavelength or a hydrogen atom sitting in its ground  
state.  In these cases, 
$\Td \y 10^{-19} s$ and $\Td \y 10^{-14} s$,  
respectively.  Thus for these systems spacetime coarse
grainings can extend only over very short time scales if they
are to be approximated by instantaneous alternatives. 
At a  
much larger scale, take the center of mass of a 
grain of dust with a diameter of about one micron and a  
corresponding mass on the order of $10^{-15} kg$ localized to its
dimension.
The resulting time scale for decoherence is then on  
the order of a year.  For any 
macroscopic particle (for example with a mass of one gram and  
a size on the order of a centimeter) $\Td$ is 
enormously greater than the age of the universe.   
For such systems the time scale $\Td$ over which this kind of
spacetime alternatives may extend while still automatically decohering
is much longer than characteristic dynamical time scales $T_{\rm dynamical}$. 

These results suggest that, for quantities such as the center of mass
position of a body characterized by typical macroscopic masses and
uncertainties, there are a class of spacetime coarse grainings
extending over a time $T \ll T_{\rm dynamical} \ll \Td$  to which
instantaneous alternatives are an excellent approximation both with
respect to decoherence and with respect to probabilities. For systems
of small mass, these results suggest that the regime of validity of
such approximations may be more
limited. In particular, in realistic measurement situations of light 
systems which extend over time, it may be necessary to take the details
of the experimental arrangement into account,  so that the alternatives
describing the outcome of the measurement refer to the alternative
configurations of the apparatus rather than the system being measured 
if they are to be well approximated
by instantaneous alternatives. It would be desirable to have more
detailed and realistic models to confirm this.

\vskip 1cm

\centerline{\bf Acknowledgments}
The authors are grateful to David Craig for much advice on the
evaluation of the integrals in this paper. This work was
supported under NSF grants PHY94-07065 and PHY94-07194.  
The work of R.M. was supported by a UCSB College of Creative Studies Summer Undergraduate Research Fellowship.
  
\listrefs

\end